\documentstyle[twoside,fleqn,espcrc2,rotate]{article}

\input{psfig}

\newcommand{\rf}[4]{{\em {#1}} {\bf #2}, #3 (#4)}

\newcommand{\pr}{Phys.\ Rev.\ }

\newcommand{\pl}{Phys.\ Lett.\ }

\newcommand{\np}{Nucl.\ Phys.\ }

\def\be{\begin{equation}}
\def\ee{\end{equation}}
\def\bea{\begin{eqnarray}}
\def\eea{\end{eqnarray}}
\def\Tr{{\rm Tr}\,}
\def\muhat{\hat{\mu}}
\def\qhat{\hat{q}}
\def\bra{\langle}
\def\ket{\rangle}

\begin{document}

\title{The structure of the gluon propagator\thanks{Talk presented by J. I. Skullerud}}

\author{D. B. Leinweber\address{CSSM and the
Department of Physics and Mathematical Physics,\\
The University of Adelaide, SA 5005, Australia}, 
C. Parrinello\address{Department of Mathematical Sciences, University
of Liverpool,\\ Liverpool L69 3BX, England}\thanks{UKQCD
Collaboration}, 
\addtocounter{footnote}{-1}
\addtocounter{address}{-1}
J. I. Skullerud\addressmark\footnotemark 
and A. G. Williams\addressmark}

\begin{abstract}

The gluon propagator has been calculated for quenched QCD in the
Landau gauge at $\beta=6.0$ for volumes $16^3\times 48$ and
$32^3\times 64$, and at $\beta=6.2$ for volume $24^3\times 48$.  The
large volume and different lattice spacings allow us to identify and
minimise finite volume and finite lattice spacing artefacts.  We also
study the tensor structure of the gluon propagator, confirming that it
obeys the lattice Landau gauge condition.

\end{abstract}

\maketitle

\section{Introduction}
\label{sec:intro}

The infrared behaviour of the gluon propagator is important for an
understanding of confinement.  Previous conjectures range from a
strong divergence~\cite{bp} to a propagator that vanishes
in the infrared~\cite{gribov,stingl}.
Lattice QCD should be able to resolve this issue by
first-principles, model-independent calculations.  However, previous lattice
studies~\cite{bps,mms} have been inconclusive 
since they have not been able to access sufficiently low momenta.  
Here we will report results using an asymmetric lattice with a spatial length of 3.3~fm  This gives us
access to momenta as small as 400~MeV.

\section{Lattice formalism}
\label{sec:def}

We use the `symmetric' definition of the gluon field, given
by $U_\mu(x) =  \exp(ig_0aA_\mu(x+\muhat/2))$.
This gives the gluon field in momentum space,
\be
A_\mu(\qhat) = 
 \frac{e^{-i\qhat_{\mu}a/2}}{2ig_0a}\left[B_\mu(\qhat)
-\frac{1}{3}\Tr B_\mu(\qhat)\right]
\ee
where  $\qhat$ denotes
the discrete momenta $\qhat_\mu = 2 \pi n_\mu/(a L_\mu)$,
$B_\mu(\qhat)\equiv U_\mu(\qhat)-U^{\dagger}_\mu(-\qhat)$, and
$U_\mu(\qhat)\equiv\sum_x e^{-i\qhat x}U_\mu(x)$.  An
alternative, `asymmetric' definition of the gluon field can be
provided by $U_\mu(x) = \exp(ig_0aA'_\mu(x))$.  In momentum space,
this differs from $A_\mu(x)$ by a factor $\exp(i\qhat_\mu a/2)$.

The gluon propagator $D^{ab}_{\mu\nu}(\qhat)$ is defined as
\be
D^{ab}_{\mu\nu}(\qhat) = \bra A^a_\mu(\qhat)
A^b_\nu(-\qhat) \ket\,/\,V \, ,
\ee
where $A_\mu(\qhat)\equiv t^a A_{\mu}^a(\qhat)$.  In the continuum
Landau gauge, the propagator has the structure
\be
D_{\mu\nu}^{ab}(q) =
\delta^{ab}(\delta_{\mu\nu}-\frac{q_{\mu}q_{\nu}}{q^2})D(q^2)
\, ,
\label{eq:landau-prop}
\ee
At tree level, $D(q^2)$ will have the form $D^{(0)}(q^2) = 1/q^2$.
On the lattice, this becomes
$D^{(0)}(\qhat) = 
a^2/(4\sum_{\mu}\sin^2(\qhat_{\mu}a/2))$.
Since QCD is asymptotically free, we expect that up to logarithmic
corrections, $q^2 D(q^2) \to 1$ in the ultraviolet.  Hence we define
the new momentum variable $q$ by
$q_\mu \equiv (2/a)\sin(\qhat_\mu a/2)$, and use this throughout

\begin{table}[tb]
\caption{Simulation parameters.  The lattice spacing is taken from the
string tension \protect\cite{bs}.}
\label{tab:sim-params}
\begin{center}
\leavevmode
\begin{tabular}{lcccr}
\hline
Name &$\beta$ &$a^{-1}$ (GeV) &Volume &$N_{\rm conf}$ \\
\hline
Small    &6.0  &1.885   &$16^3\times 48$ &125 \\
Large    &6.0  &1.885   &$32^3\times 64$ &75  \\
Fine     &6.2  &2.63    &$24^3\times 48$ &223 \\
\hline
\end{tabular}
\end{center}
\end{table}

We have analysed three lattices, with different values for the volume
and lattice spacing.  The details are given in
table~\ref{tab:sim-params}.  All the configurations have been fixed to
Landau gauge with an accuracy $\langle(\partial_\mu A_\mu)^2\rangle <
10^{-12}$.

\section{Tensor structure}
\label{sec:tensor}

By studying the tensor structure of the gluon propagator, we may be
able to determine how well the Landau gauge condition is satisfied,
and also discover violations of continuum rotational invariance.

The continuum tensor structure (\ref{eq:landau-prop}) follows from the
condition $q_\mu A_\mu=0$.  This translates directly to the lattice
provided we use the symmetric definition of the gluon field.  If we
use the asymmetric definition, we will instead obtain the condition
$\sum_\mu(i\sin\qhat_\mu+\cos\qhat_\mu-1)A'_\mu(\qhat) = 0$.

The tensor structure may be measured directly by taking the ratios of
different components of $D_{\mu\nu}(q)$ for the same value of $q$.
The results for the small lattice are summarised in table
\ref{tab:tensor-small}, and compared to what one would expect from
(\ref{eq:landau-prop}), and to what one would obtain by replacing $q$
with $\qhat$ in (\ref{eq:landau-prop}).  The results are similar for
the two other lattices.
It is clear from table \ref{tab:tensor-small} that our numerical data
are consistent with the expectation from (\ref{eq:landau-prop}).  We
can also see that in general, the asymmetric definition $A'$ of the
gluon field gives results which are inconsistent with this form.

\begin{table*}[htbp]
\caption{Tensor structure for the small lattice. $\qhat$ is in units
of $2\pi/L_s$, where $L_s$ is the spatial length of the lattice.  The
theoretical predictions are the values for the ratios one obtains from
(\protect\ref{eq:landau-prop}), and from
(\protect\ref{eq:landau-prop}) with $q\to\qhat$.  The numbers in
brackets are the statistical uncertainties in the last digit(s).
Where no error is quoted, the statistical uncertainty is less than
$10^{-6}$.}
\label{tab:tensor-small}

\begin{tabular*}{\textwidth}{cc@{\extracolsep{\fill}}rrrr} 
\hline
 & & \multicolumn{2}{c}{Theoretical prediction} &
 \multicolumn{2}{c}{This simulation} \\ \cline{3-4}\cline{5-6}
{\bf $\qhat$} & Components & Using $\qhat$ & Using $q$ &
 Using $A$ &  Using $A'$ \\ \hline
[2,1,0,0]
& (1,1)/(1,2) &  -0.5 &  -0.509796 &  -0.509796 &  -0.519783 \\
& (1,1)/(2,2) &  0.25 &   0.259892 &   0.259892 & 0.259892  \\
& (1,1)/(3,3) &   0.2 &   0.206281 &   0.204(8) & 0.204(8)  \\
& (1,2)/(2,2) &  -0.5 &  -0.509796 &  -0.509796 & -0.5  \\ \hline

[4,1,0,0]
& (1,1)/(1,2) & -0.25 &  -0.275899 &  -0.275899  & -0.331821 \\
& (1,1)/(3,3) & 0.05882 &   0.070736 &  0.076(3) & 0.076(3) \\
& (1,2)/(2,2) & -0.25 &  -0.275899 &  -0.275899  & -0.229402 \\ \hline

[4,2,1,0]
& (1,1)/(1,2) & -0.625 &   -0.681848 &  -0.678(9) & -0.743(10) \\
& (1,1)/(2,3) &   -2.5 &    -2.47137 &    -2.3(4) &    -2.5(5) \\
& (1,2)/(2,2) & -0.4706 &  -0.502914 &  -0.500(6) &  -0.456(6) \\ \hline

\end{tabular*}
\end{table*}

\section{Finite size effects and anisotropies}
\label{sec:anisotropy}

In the following, we are particularly interested in the deviation of
the gluon propagator from the tree level form.  We will therefore
factor out the tree level behaviour and plot $q^2 D(q^2)$ rather than
$D(q^2)$ itself.

\begin{figure}[t]
\begin{center}
\leavevmode
\rotate[l]{\psfig{figure=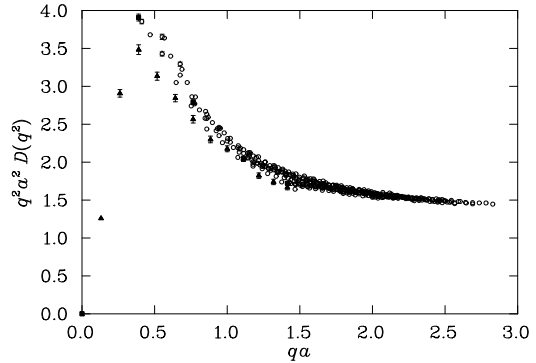,height=2.7in}}
\end{center}
\vspace{-1.2cm}
\caption{The gluon propagator multiplied by $q^2$ as a function of
$q$ for the small lattice.  The filled
triangles denote momenta directed along the time axis, while the
filled squares denote momenta directed along one of the spatial axes.}
\label{fig:cpt-small}
\vspace{-15pt}
\end{figure}

\begin{figure}[hbt]
\begin{center}
\leavevmode
\rotate[l]{\psfig{figure=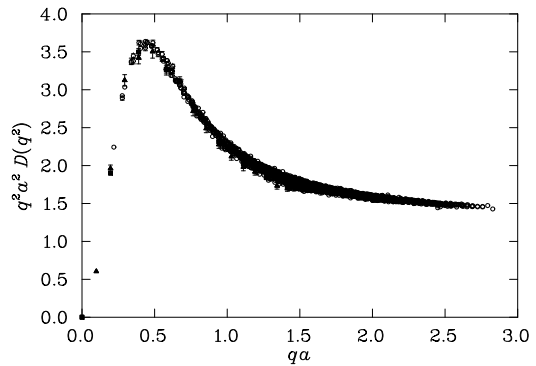,height=2.7in}}
\end{center}
\vspace{-1.2cm}
\caption{The gluon propagator multiplied by $q^2$ as a function of $q$
for the large lattice.  The symbols are as in
fig.~\protect\ref{fig:cpt-small}.}
\label{fig:cpt-large}
\vspace{-15pt}
\end{figure}

Fig.~\ref{fig:cpt-small} shows the gluon propagator on the small
lattice as a function of $qa$.  For low momentum values on the small
lattice, there are large discrepancies due to finite volume effects
between points representing momenta along the time axis and those
representing momenta along the spatial axes.  These discrepancies are
absent from the data from the large lattice, shown in
fig.~\ref{fig:cpt-large}.  This indicates that finite volume effects
here are under control.

\begin{figure}[hbt]
\vspace{-5pt}
\begin{center}
\leavevmode
\rotate[l]{\psfig{figure=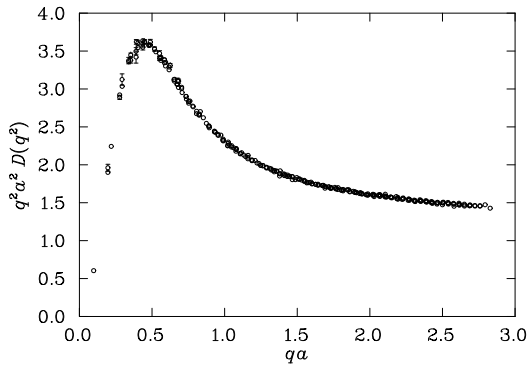,height=2.7in}}
\end{center}
\vspace{-1cm}
\caption{The gluon propagator multiplied by $q^2$ as a function of $q$
for the large lattice, after the cylindrical cut.}
\label{fig:cut-large}
\end{figure}

However, at higher momenta, there are anisotropies which remain for
the large lattice data, and which are of approximately the same
magnitude for the two lattices.  These anisotropies are considerably
reduced on the fine lattice,
indicating that they arise from finite lattice spacing errors.  In
order to eliminate these anisotropies, we select momenta lying within
a cylinder of radius $\Delta\qhat a = 2\times 2\pi/32$ along the
4-dimensional diagonals.  The result of this cut on the large lattice
is shown in fig.~\ref{fig:cut-large}.  A more detailed discussion of
these cuts can be found in \cite{letter}.

\section{Conclusions}
\label{sec:discuss}

We have evaluated the gluon propagator on three different lattices.
The tensor structure has been analysed and shown to agree with the
continuum Landau gauge form.  By studying the anisotropies in the
data, we have been able to conclude that finite volume effects are under
control on the largest of our lattices.

A clear turnover in the behaviour of $q^2 D(q^2)$ has been observed at
$q \sim 1$GeV, indicating that the gluon propagator diverges less
rapidly than $1/q^2$ in the infrared, and may be infrared finite.  An
analysis of scaling and the functional behaviour of these results is
presented in~\cite{dbl}.

\section*{Acknowledgments} 

The numerical work was mainly performed on a Cray T3D at EPCC,
University of Edinburgh, using UKQCD Collaboration time under PPARC
Grant GR/K41663.  Financial support from the Australian Research
Council is gratefully acknowledged.


\begin{thebibliography}{99}

\bibitem{bp} N.~Brown and M.R.~Pennington, \rf{\pr}{D 39}{2723}{1989}

\bibitem{gribov} V.N.~Gribov, \rf{\np}{B 139}{19}{1978}

\bibitem{stingl} M.~Stingl, \rf{\pr}{D 34}{3863}{1986}; 
\rf{\pr}{D 36}{651}{1987} 

\bibitem{bps} C.~Bernard, C.~Parrinello, A.~Soni,
\rf{\pr}{D49}{1585}{1994} 

\bibitem{mms} P.~Marenzoni, G.~Martinelli, N.~Stella, 
\rf{\np}{B 455}{339}{1995}; P.~Marenzoni {\em et al}, 
\rf{\pl}{B 318}{511}{1993}

\bibitem{bs} G.S.~Bali and K.~Schilling, \rf{\pr}{D 47}{661}{1993}

\bibitem{letter} D.B.~Leinweber, C.~Parrinello, J.I.~Skullerud,
A.G.~Williams, \rf{\pr}{D 58}{031501}{1998}

\bibitem{dbl} D.B.~Leinweber, C.~Parrinello,
J.I.~Skullerud, A.G.~Williams: {\em Modelling the gluon
propagator}, these proceedings.

\end{thebibliography}
\end{document}